# Universality of time evolution for the unsteady dendrite

L.M. Martyushev, P.S. Terentiev
Institute of Industrial Ecology, Russian Academy of Sciences, 20 S. Kovalevskaya St.,
620219 Ekaterinburg, RUSSIA
Ural Federal University, 19 Mira St., Ekaterinburg, 620002, RUSSIA
LeonidMartyushev@gmail.com

Unsteady growth of ammonium chloride dendrites during crystallization from an aqueous solution in a thin capillary is experimentally investigated. Dependency of the crystal area $S$ on the time $t$ for various sectors located along a primary branch and sidebranches is measured. A hypothesis of the same ratio between the area change and the area itself ($S'(t)/S(t)$) for different but simultaneously growing parts of an unsteady dendrite is advanced and confirmed. This conclusion allows proposing a curve for describing the evolution of the dendrite area (or its part) of the form $S(t) = const \cdot t^a \cdot \exp(-b \cdot t)$, where $a$ and $b$ are the parameters, whose values are determined in the paper. The nondimensionalization of S(t) and $S'(t)/S(t)$ (using the full dendrite growth time) results in simple one-parameter functions depending on a single parameter $a$ (which is presumably associated only with physical and chemical characteristics of the crystallized system and, in our case, equals $1.7 \pm 0.2$).

## I. INTODUCTION

The interest in the investigation of the dendrite growth during crystallization arose long ago [1] and remains until the present time [2-9]. This is primarily connected with the fact that these structures often form during the solidification of melts and determine many properties of metal ingots [2-4]. In addition, profound interest in the dendrite growth is shown by the researchers studying not the material science but various problems of the physics of non-equilibrium processes [5-9]. This is due to the dendrites that form in the multiple so-called non-equilibrium (or dissipative) processes commonly observed in nature. There are several examples of the formation of dendrite-like (also sometimes referred to as fractal) structures, such as lightning discharges, growing trees, development of transport infrastructure [9,10]. Nature often chooses these structures when it is necessary to dissipate, remove the non-equilibrium state (supercooling or supersaturation, change of electric potential, etc.) generated for some reason. The researchers therefore study the dendrite crystallization as an example on which basis the general regularities of non-equilibrium processes can be understood. The example is remarkable because it is easy in the experimental investigation, the underlying physical processes are well studied, and the describing equations are relatively few and simple [5-7].

---

[1] Thus, as far back as 1611, Kepler dedicated his paper to an ice dendrite, i.e. a snowflake [1].

Many papers dealt with the regularities of the dendrite development during crystallization (see, for example, reviews [5-8]). The relation between the external shape (or the morphology) and the growth rate (or the kinetics) of the dendrite is one of the most studied fields. At present, it is considered that the dendrite has a rigorously defined (close to parabolic) shape and a precise rate at certain supercooling/supersaturation. When increasing supercooling/supersaturation, the dendrite increases its growth rate $V$ and becomes sharper (the dendrite tip radius $R$ decreases). It is further shown that the quantity $VR^2$ in the first approximation does not depend on the supercooling/supersaturation but is determined by the reference parameters of the crystallized system. The given results received both experimental and theoretical confirmation (see, for example, [5-8]) and represent important achievements, however many questions still remain open. First of all, the mentioned facts relate to the steady dendrite growth. However, it may turn out to be very difficult to precisely detect this dendrite growth stage (especially for the dendrites growing under natural uncontrolled conditions rather than laboratory and controlled conditions when supersaturation/supercooling is artificially maintained constant). Indeed, the dendrite crystal has three growth stages: the origination, more or less steady growth, and the stage of the rate decrease and the stopping (when supercooling/supersaturation is removed as a result of crystallization). Obviously, the constancy of $VR^2$ will not hold during the whole interval of the dendrite growth[2]. The inapplicability of the obtained results to sidebranches is the second disadvantage. Indeed, the sidebranches grow at slightly lower supercooling/supersaturation values and show smaller rate than that of the primary branch. According to the criterion of the constancy of $VR^2$, they are to have blunter tips. However, in actual practice, their tips typically have a considerably smaller radius of curvature compared to the primary branch. As a result, the researchers aiming to confirm/verify the constancy of $VR^2$ are forced to sufficiently artificially divide the dendrite into the primary branch (growing in conformity with the theory) and the sidebranches (not conforming to the theory). This leads to some loss of universality and practical significance of the results[3].

---

[2] When decreasing supercooling/supersaturation, $R$ increases tending to some constant value and the rate is reduced to zero. It should be noted that a number of recent experiments (see papers [8, 11-13]) called in question the independency of $VR^2$ from supersaturation even in the case of steady growth.
[3] Indeed, when we focus the microscope on an arbitrary spot of the solution and find a dendrite, we cannot foretell what we shall see. It may be a primary branch with secondary branches, or it may be a quite big and detached (as a result of the increase of side-branch intervals due to the period doubling) secondary branch with the forming tertiary branches. In

Let us make an intermediate summary. Dendrite crystallization is a very common phenomenon in nature. The parameter $VR^2$ cannot pretend to the role of a universal one as it is inapplicable to an arbitrary (unsteady) interval of time evolution of the dendrite and fails to describe the morphokinetics of the dendrite sidebranches (and is only applicable to the region near the tip of the primary branch). Is there such universal parameter at all, and if there is one, what is it? Let us provide two relatively independent lines of non-rigorous reasoning indicating that such parameter evidently exists.

1. As it was mentioned, the dendrite growth often accompanies a non-equilibrium process. The so-called maximum entropy production principle (MEPP) has received wide recognition for non-equilibrium processes of different nature in the past few decades (see, for example, reviews [14-17]). This principle states that a system responds to an external influence such as to maximize the entropy production[4]. For the crystal growing in non-equilibrium manner, this principle has been successfully applied many times [16-25]. The fact that the simultaneous development of different crystal patterns (the so-called co-existence of different morphological phases) occurs subject to the equality of entropy productions of these growing crystal patterns is one of the consequences of the principle[5]. Let us consider a dendrite as a set of co-existing primary and sidebranches that grow simultaneously tending to remove the initial non-equilibrium state (supersaturation or supercooling). The entropy production of a unit volume crystal in the first approximation is proportional to the crystal volume increment in the time ($t$) divided by the volume [17-22]. If we consider that the crystal grows in a quasi-two-dimensional manner, than this quantity will equal $S'(t)/S(t)$, where $S(t)$ is the area of some element of the crystal branch near the crystallization front.

Thus, according to MEPP, the local entropy productions for the primary branch and the sidebranches of the dendrite growing in a non-equilibrium manner are equal at some point in time and specifically the value of $S'(t)/S(t)$ should be approximately the same for different quasi-two-dimensional branches of the dendrite[6].

2. The dendrite represents a certain ordered set of branches with parabola-like shapes (for

---

the second case, the researcher can take the secondary branch as a primary one even when using a different spatial scale (when zooming in with the microscope).
[4] Furthermore, the principle is applied locally, therefore the entropy production density can be mentioned.
[5] It is obvious that the equality of the entropy productions of one (for example, unit) volume is meant.
[6] Here, approximateness is mentioned, as the accurate calculation of the local entropy production requires further consideration of the change of the chemical potential near the crystallization front and more proper treatment of the volume near the crystallization front where, in fact, dissipation occurs.

simplicity and definiteness, we shall consider below the two-dimensional case and the crystallization from a solution). Let us consider a single parabola and select a certain sector therein with an arbitrary angle α (for example, p1 in Fig. 1). The equation of the parabola moving in the 0Y axis direction at the rate V(t) and with the tip radius R(t) has the following form:

$$y(x,t) = -x^2/(2 \cdot R(t)) + \int_0^t V(t)dt.$$

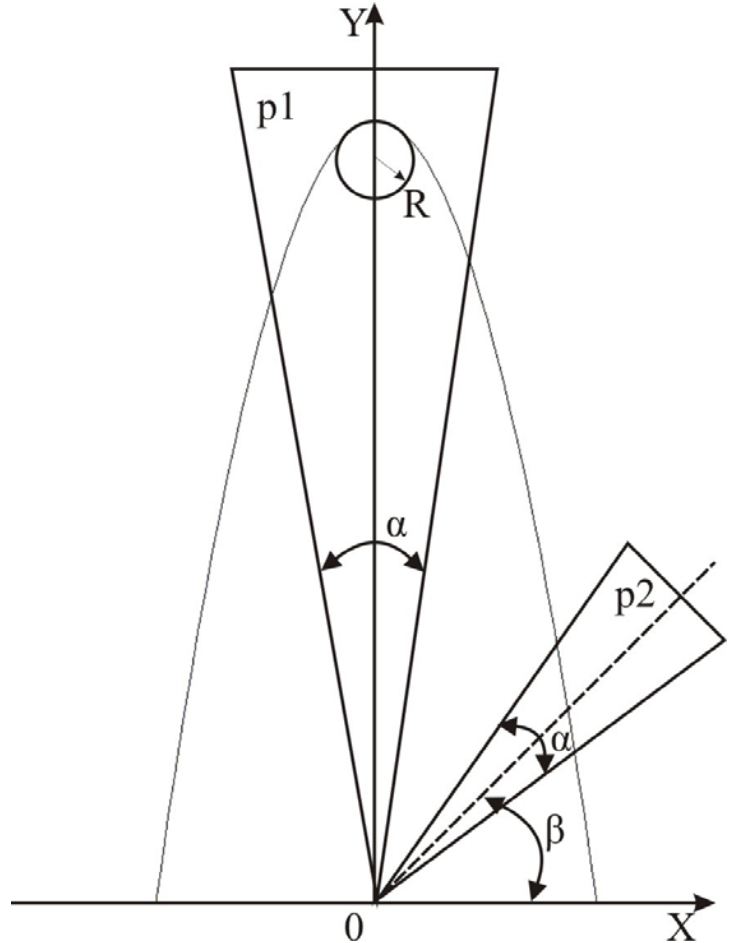

FIG. 1. Parabola with different sectors, where the area is calculated.

As a result, the parabola area will be written as $S(t) = 2\int_0^{d(t)} y(x,t)dx - y(d(t),t) \cdot d(t)$,

where $d(t)$ is the positive solution of the equation $x \cdot ctg(\alpha/2) = -x^2/(2 \cdot R(t)) + \int_0^t V(t)dt$. As it is shown above, the specific change of crystal area (which can also be referred to as the increment per unit of crystal area, the normalized area increment, or the relative increment) $S'(t)/S(t)$ has an interesting property. Let us calculate this quantity for the two simplest and most widespread crystallization models.

i) We shall assume that $V(t)$ and $R(t)$ are just some constants (related to supersaturation, diffusivity, etc., but independent of time), i.e. the steady dendrite growth occurs. In this case, for small angles α: $S'(t)/S(t)=2/t$. Thus, this quantity proves to be independent of both supersaturation and other crystal characteristics.

ii) We shall assume that the parabolic crystal grows under quasi-steady diffusion-limited conditions. As it is known, in this case $V(t)=\Omega/\sqrt{t}$ and $R(t)=\Xi\sqrt{t}$, where $\Omega, \Xi$ are some constants related to supersaturation, and physical and chemical parameters of the considered crystallized system [26, 27]. Then, in this case $S'(t)/S(t)=1/t$. Moreover, this ratio, as opposed to the previous case, proves to be true for any angle α.

Thus, the simplest models considered indicate the universality of the quantity $S'(t)/S(t)$. This quantity has power-law dependence only on time during which the parabolic[7] dendrite grows in the sector selected along its direction of growth. Let us note that if the selected sector is at some angle β to the direction of growth (Fig. 1), then the quantity $S'(t)/S(t)$ becomes dependent on many quantities, and there is no such simple and universal power-law dependence on time. According to the analysis results, the numerical factor multiplying time (in the considered cases, it is 0.5 or 1) depends not only on the dependence of $V$ and $R$ on time, but also on the sector shape, where the dendrite is observed (in particular, the numerical factors turn out to be twice smaller for the rectangular sectors), and, as it is easy to verify, on the spatial dimension of problem.

The real dendrite consists of a primary branch and sidebranches with parabola-like shapes. Will the experiment actually confirm the constancy of the quantity $S'(t)/S(t)$ for various dendrite branches growing simultaneously? Will this quantity depend on the angles α and β of the sectors? How close, in the case of real unsteady dendrites, will the quantity $S'(t)/S(t)$ be to the dependence, which is given above for the simplest models of the steady and the quasi-steady growth. The search for the answers to these questions forms the objective of the present study.

---

[7] The parabolicity of the dendrite front is not of fundamental importance, it can be easily proved that the similar conclusions will be true also, for example, in the case of rounded or rectilinear shapes of the crystallization front.

## II. EXPERIMENTAL PROCEDURE

### A. Experimental setup

The aqueous solution of ammonium chloride ($NH_4Cl$), whose crystallizations have been repeatedly studied before, was selected as a dendrite growth system [28-36].

The prepared $NH_4Cl$ solution with the concentrations of 43.6 (g/100g $H_2O$), which corresponds to the saturation temperature of 35°C, was placed between two glasses of 18 x 18 x 0.2 mm. The solution thickness between the glasses did not exceed 0.05 mm. Since the diffusion length was not less than $2D/v \approx 0.5$ mm (where $v$, $D$ are the growth rate and the dendrite diffusivity, respectively ($D=2.6 \times 10^{-5} cm^2/s$ [32])), the cell prepared in such a manner can be considered as a quasi-two-dimensional cell. The prepared cell was held at the temperature exceeding the saturation temperature by 5°C to homogenize the solution in the cell and remove the possible crystal nuclei. The waiting time was not less than 10 minutes. Then the cell was placed on a heavy objective table of the BIOLAR PI microscope at the temperature of 20°C. Due to the thin coverglasses and the small thickness of the solution between them, the cell took the temperature of 20°C within 2-3 seconds. The video recording of freely growing dendrites was started after 10-15 seconds. The observation (in the majority of cases, with 110-fold magnification) and the video recording applied only to single freely-growing dendrites separated from the neighboring dendrites by at least the diffusion length. In the vast majority of cases (94%), the dendrite growth with the secondary branches oriented in the direction of <100> was observed[8]. No tertiary branches were observed. In total, 96 samples were studied. The experiments showed less (Fig. 2) and more (Fig. 3) symmetrical dendrites. The share of the first ones was slightly bigger (around 65 percent).

The recording was carried out using a digital video camera with the resolution of 720x576, the size of a pixel at the selected zoom was 0.00068 mm.

### B. Image processing and measurement errors

The crystal area (directly proportional to the crystal mass, because the dendrite growth occurred under the quasi-two-dimensional conditions) was measured as follows:

1. A video clip of the dendrite growth was divided into separate frames with the frequency of at least 1 frame/sec. These frames were converted to the grayscale. Then, each frame was individually processed. For this purpose, a special software tool was written in the MATLAB (Image Processing Toolbox) environment.

---

[8] The so-called irregular patterns [30] were observed in 4% of the cases. A separate paper will be dedicated to the discussion of the kinetics of these crystals.

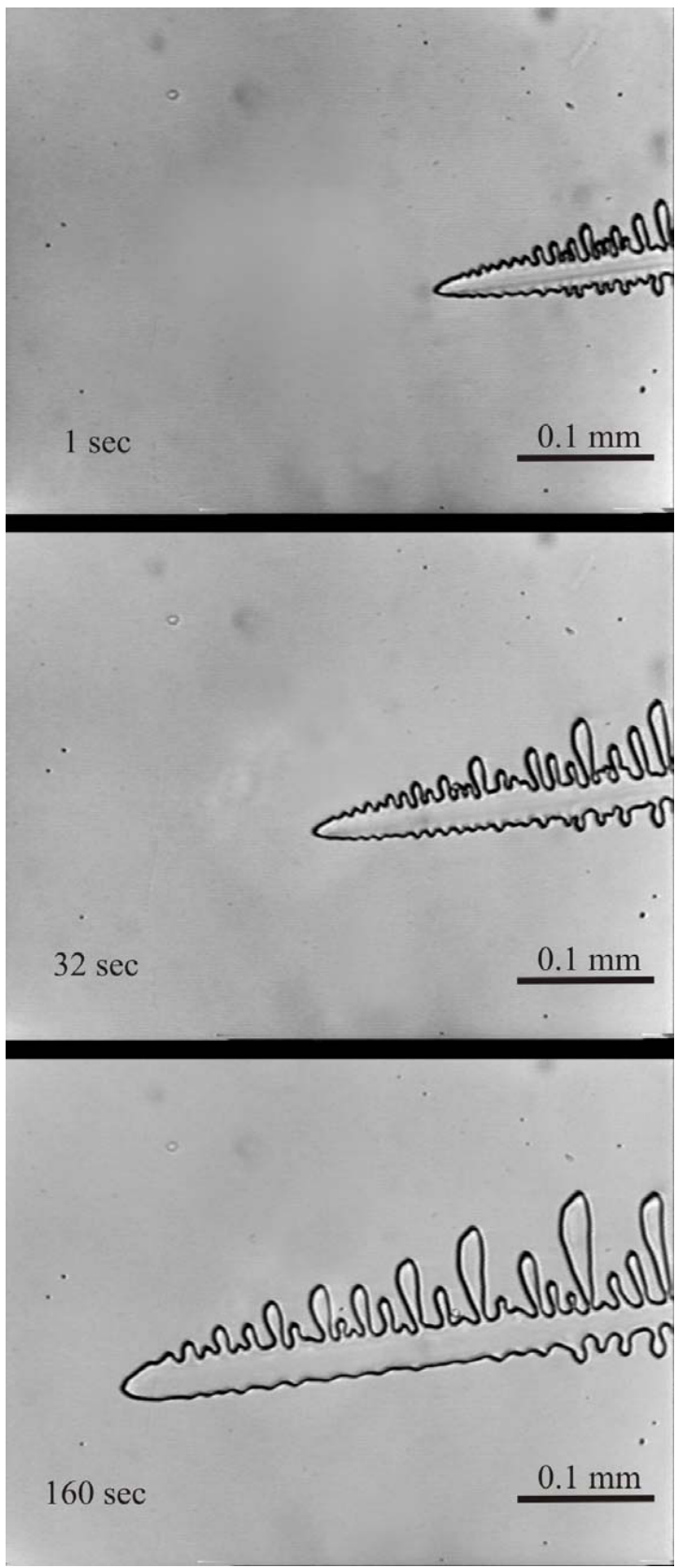

FIG. 2. Frames of an asymmetric dendrite of ammonium chloride growing from the aqueous solution.

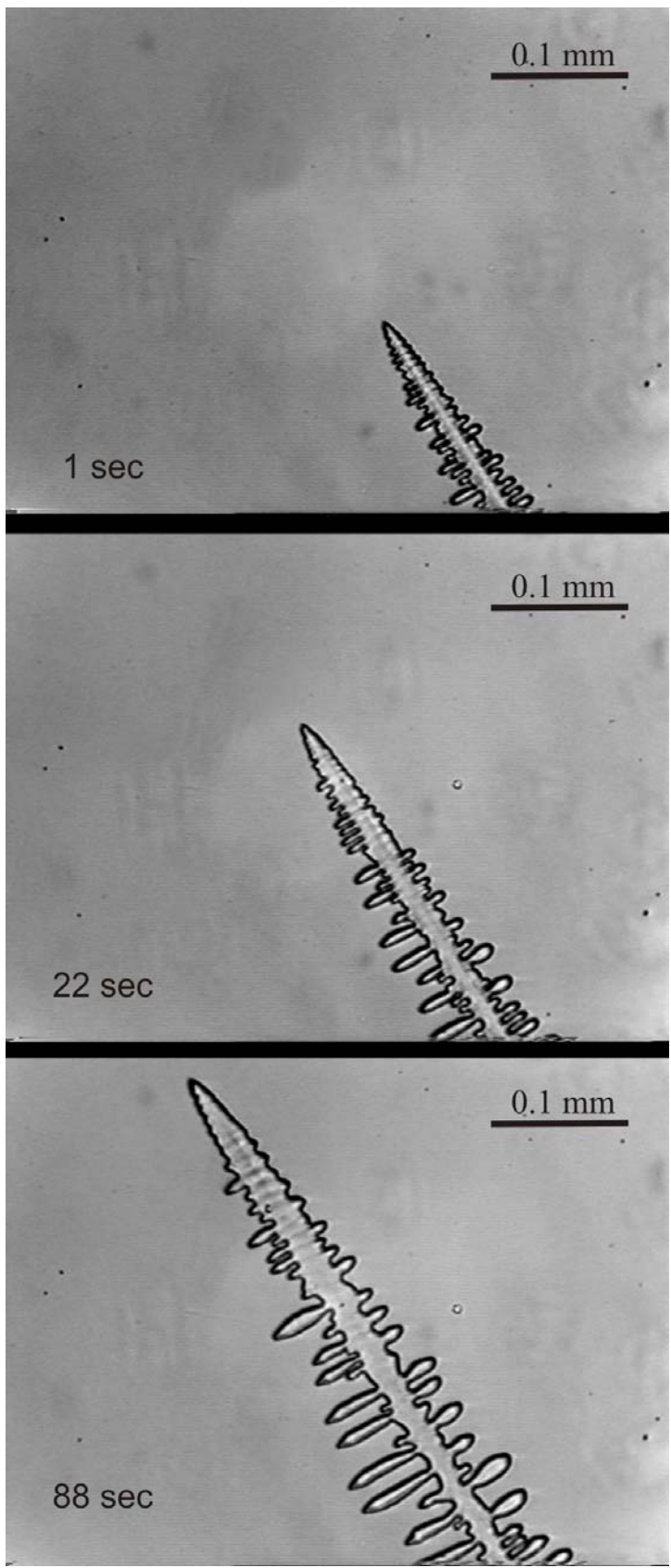

FIG. 3. Frames of a symmetric dendrite of ammonium chloride growing from the aqueous solution.

2. The detection of the crystal contour in each frame is the most complicated stage of image processing. This is due to the fact that the processing requires automation, as the number of frames for each video clip can amount to approximately one thousand. Two standard methods were used: the method of morphological erosion and the method of brightness histogram adjustment [37]. The first method is based on the detection of boundaries from the brightness gradient difference. This results in multiple unconnected fragments of the crystal perimeter located along its apparent boundary. The fragments form a continuous line by thickening. Such a procedure leads to a slightly over-estimated value of the crystal area due to the boundary thickening. The second method presupposes the histogram adjustment (brightness redistribution). The crystal object in the picture usually has a specific brightness interval not covering the whole possible range (from 0 to 255). Pixels of the picture, whose brightness values are beyond the interval, are assigned values of the brightness interval boundary. Such manipulation results in the noise reduction and the improvement of the crystal-background contrast. This enables to segment (detect) the edge pixels with the threshold filtering. Neither the first nor in the second method has the full process automation. For example, it is necessary to set the minimum brightness gradient difference, which is to be taken into account in the first method, and set the brightness interval and the threshold value in the case of the second method. However, if these parameters are fitted for several frames in different moments of the dendrite growth, then all intermediary frames can be processed automatically.

Testing of these two methods on frames using the structures with the known areas and with the video quality similar to the experimental one has shown that in the case of automatic area calculation the method based on histogram adjustment has greater accuracy and stability. This method was selected as the main one for image processing.

3. After detecting the dendrite contour, the images were binarized: the region belonging to the crystal was assigned one color (black), and the solution was assigned other color (white) (Fig. 4). The crystal area was found by simply counting the black pixels. The pixels were converted to mm using the templates with the known sizes video-recorded under the same conditions.

4. As it was mentioned above, the detection of the crystal boundary inevitably results in the error of the crystal area calculation. Obviously, the greater the ratio of the characteristic size of the crystal $R$ to the size of the pixel $d$, the smaller the error. The relative error of area calculation with the size determination accuracy of one pixel can be obtained from $\delta_R=2d/R$. As a consequence, for the crystal size of 10 pixels (and, respectively, the area of approximately 100-300 pixels) the absolute error of area determination is 20%, and for the characteristic size of 100 pixels it is only 2%. The given estimation shows that the area measurement with the accuracy of around 10% is possible only for the crystal sections with the area of approximately $10^3$ pixels. It can also be easily found that, in the considered case of the determination accuracy of one pixel, the relative error of determining the crystal area increment per unit time will approximately be $\delta_\Delta = 2d/\Delta$, where $\Delta$ is the change of the crystal size (it is assumed that the growing crystals are quite big and one can neglect the error of area determination). The error of determination of ($S'(t)/S(t)$) is $\sqrt{\delta_\Delta^2 + \delta_R^2}$. It is easy to calculate that, for a crystal with the characteristic dimension of 100 pixels and the increment of 10 pixels per unit time, the error will be at least 20% (however, the area change will be only one percent). Thus, based on the above, there is a restriction both for the sector size, where the crystal area is to be measured (it should not be very small), and for the observation time (such that the area change is not very little). Therefore, in the case of the so-called sigmoid curves (see below), the dependence of the crystal area (mass) on time, and the normalized area increment can be determined with adequate accuracy only in the middle section. Indeed, the crystal size is small (and $\delta_R$ is high) at the initial stage, and the area change is very small (and the role of $\delta_\Delta$ significantly increases) at the final stage. The above issue will be taken into account in the next section when the results of direct and indirect measurements are presented; the values will be given in the time interval, where their errors do not exceed 20%.

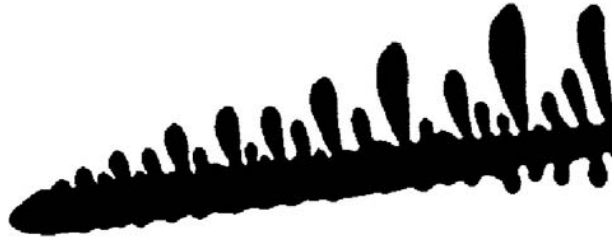

FIG. 4. Example of the crystal contour detection and the image binarization for the structure shown in Fig. 2 (160 sec).

## III. RESULTS AND DISCUSSIONS

Following the purposes of the study, the observation of the mass increment (the area increment, in the quasi-two-dimensional case considered herein) was carried out in the sectors positioned at different angles and oriented along various dendrite branches (see Fig. 5a and Fig. 6a). The measurement results of the area $S(t)$, the ratio $S'(t)/S(t)$, and the deviations of $S'(t)/S(t)$ for various dendrite sectors are given in Fig. 5 (b,c) and Fig. 6 (b,c)[9].

The following conclusions can be made based on the presented results:

1. The dependences of area on time belong to more or less pronounced sigmoid type (S-shaped curve). For different observation sectors, a significant difference is seen in the values of both $S(t)$ (sometimes by tens of times) and the crystal area increment $S'(t)$ (as an example, see Fig. 7).

2. In spite of the significant difference between $S(t)$ and $S'(t)$ for different sectors, the values of $S'(t)/S(t)$ agree in the vast majority of cases within the range of 15 percent (i.e. within the limits of the experimental error). This value has hyperbolic dependence on time. The agreement is observed not only for the small angles of 0.02 $\pi$, 0.1 $\pi$, but also for the big angles of 0.5 $\pi$ rad., $\pi$ rad., covering the whole dendrite with all sidebranches. It follows from the results given in Fig. 5c and Fig. 6c that the dependences $S'(t)/S(t)$ for the sidebranches (sectors a8-a10, s7) and the primary branches (sectors a1-a2, s1-s2) are the same within the limits of the experimental error. Let us specifically note the agreement within the error limits of the normalized increment for the dendrite parts growing in the sectors a7 and a6. These identical sectors divide the dendrite into two parts with unequal areas. Since the first sector (a7) contains the upper half of the dendrite with the developed sidebranches, and the second one contains the lower half of the dendrite with almost no branches, the agreement of the normalized mass increment is an uncommon result.

---

[9] Here and below, the results of two typical experiments out of the whole array of the obtained experimental data are presented.

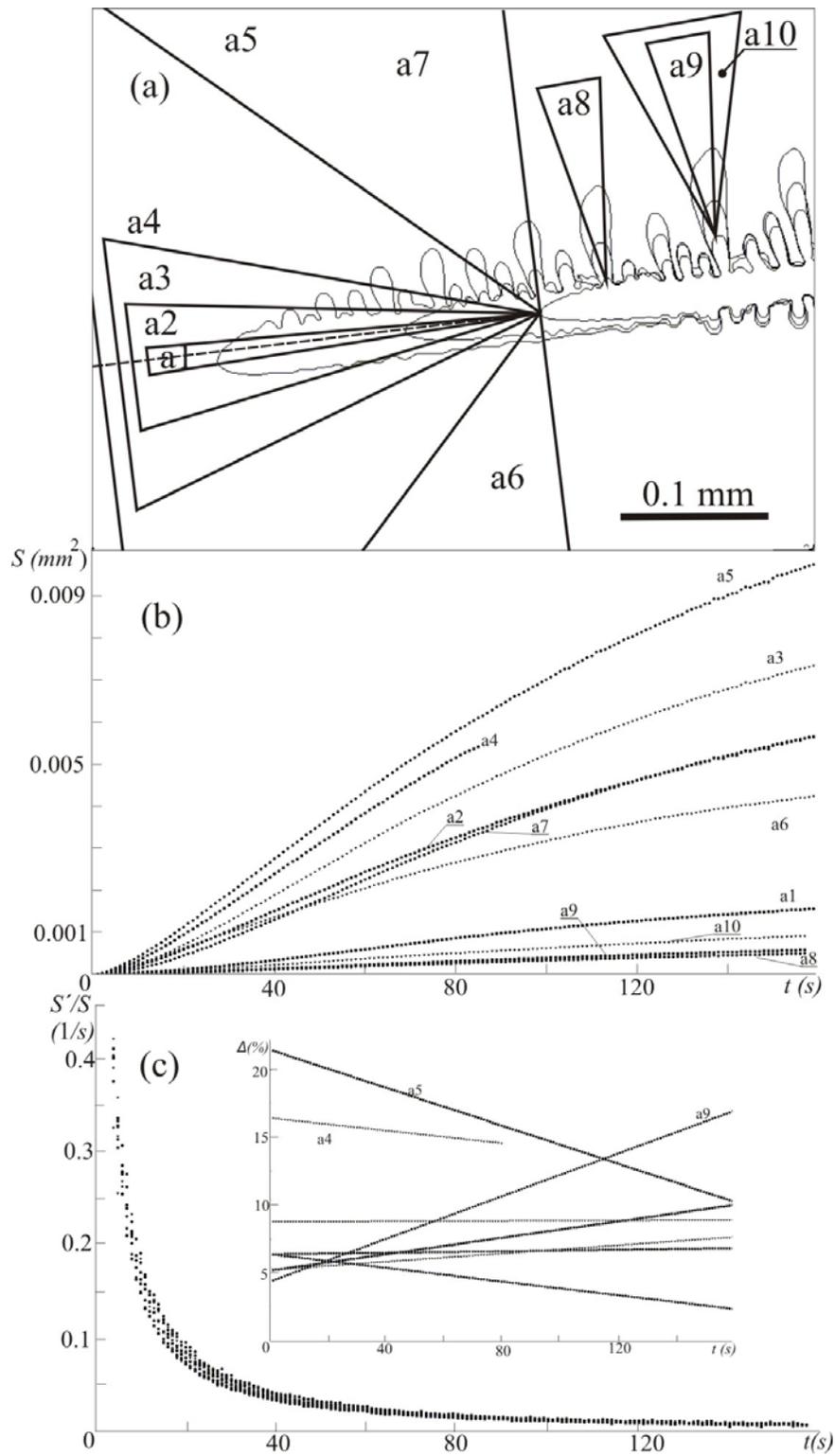

FIG. 5. Measurement results of the dendrite shown in Fig. 2. (a) Sectors used for area measurement. Sector sizes (in radians) are as follows: a1=0.02π, a2=a8=a9=0.1π, a3=a10=0.2 π, a4=a6=a7=0.5π, a5=π; (b) Values of the area $S$ relative to the dendrite growth time $t$ for different sectors; (c) Values of the normalized area increment $S'(t)/S(t)$ relative to the dendrite growth time $t$ for different sectors. The insert shows the relative difference between $S'(t)/S(t)$ of each sector and $S'(t)/S(t)$ for the sector a2.

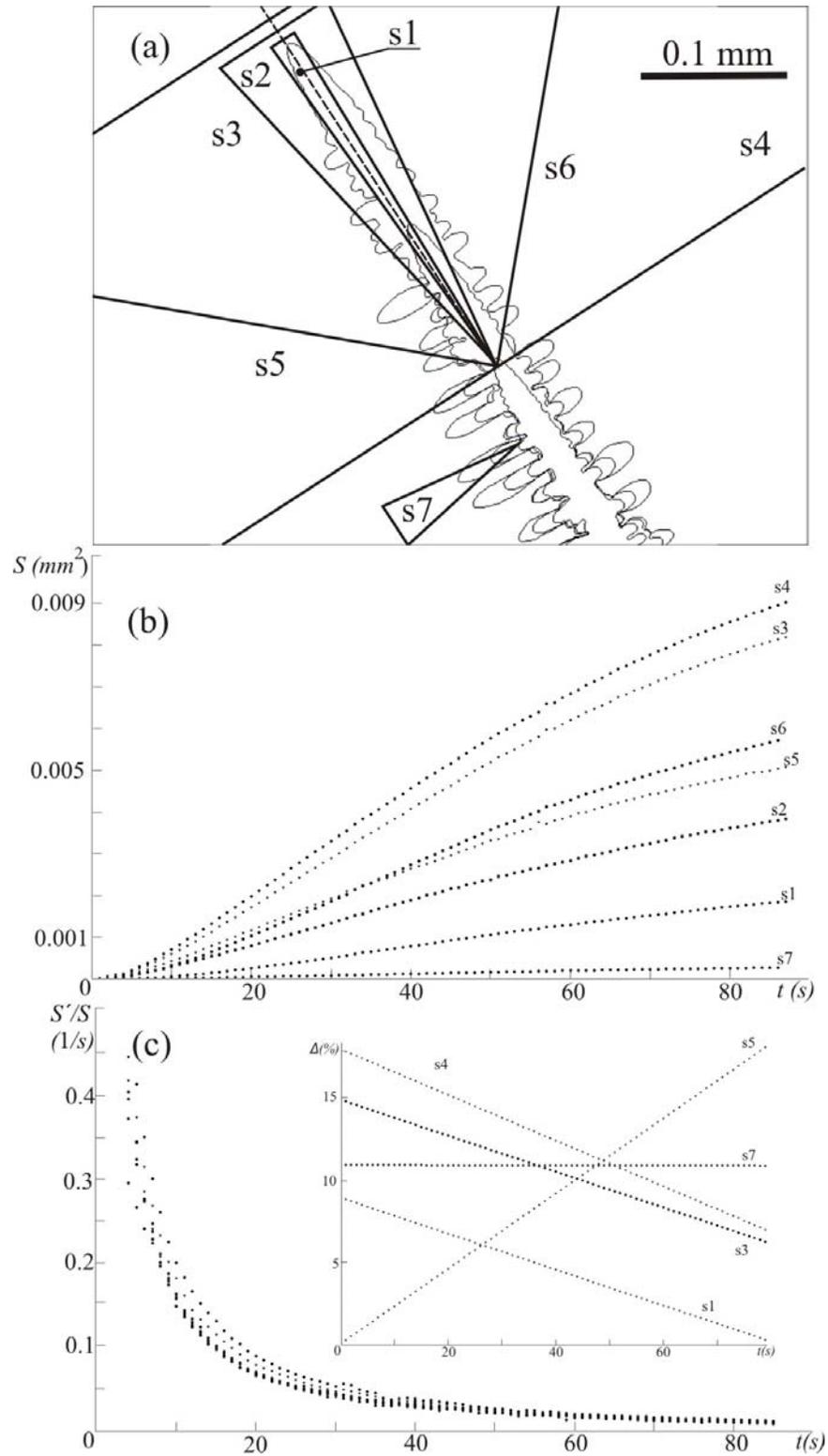

FIG. 6. Measurement results of the dendrite shown in Fig. 3. (a) Sectors used for area measurement. Sector sizes (in radians) are as follows: s1=0.02π, s2= s7=0.1π, s3=s5=s6=0.5π, s4=π; (b) Values of the area $S$ relative to the dendrite growth time $t$ for different sectors; (c) Values of the normalized area increment $S'(t)/S(t)$ relative to the dendrite growth time $t$ for different sectors. The insert shows the relative difference between $S'(t)/S(t)$ of each sector and $S'(t)/S(t)$ for the sector s2.

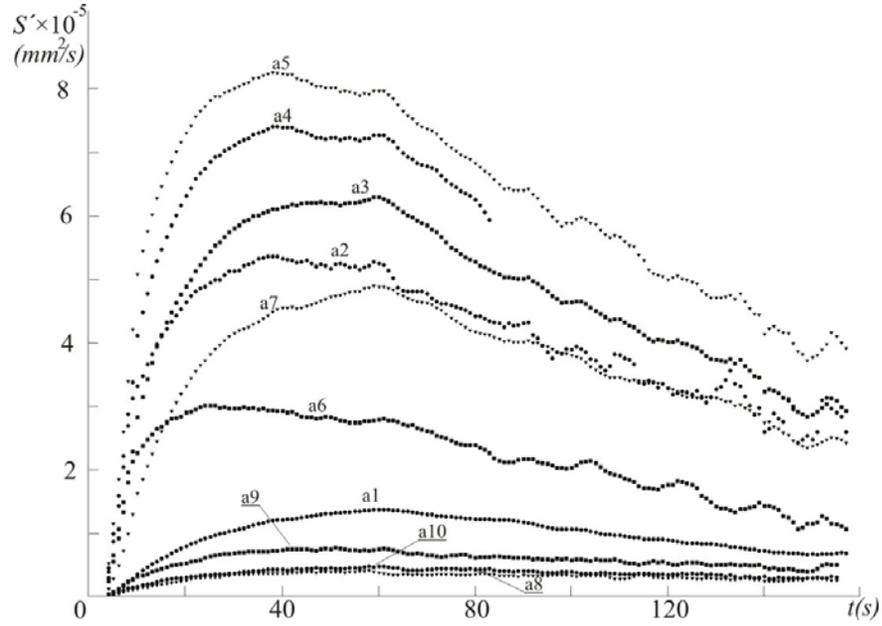

FIG. 7. Dependence of the area increment $S'(t)$ on the dendrite growth time $t$ for different sectors of the structure shown in Fig. 5.

Thus, the presented measurement results confirm the hypothesis of the universality of the normalized crystal mass increment $S'(t)/S(t)$ stated in the Introduction. However, two critical notes/issues arise:

i) Is it possible that other quantities constituted of $S'(t)$ and $S(t)$ may also be nearly equal for different sectors within the limits of the experimental error, and therefore no positive conclusion can be made based on the above results[10]? Fig. 8 shows an example of two calculations of the quantities $S'(t)/S^2(t)$ and $S'(t)/L(t)$ (where $L(t)$ is the length of the crystal boundary inside the sector) for the sectors a2 and a5. It is seen that these derived quantities for the two sectors differ by 80 to 350 percent, which enables to remove the uncertainty in question.

---

[10] The more so because, due to the sigmoid-type curves of $S(t)$, the numerator of the ratio $S'(t)/S(t)$ decreases and the denominator increases, and, as a consequence, the value under study becomes very small with time and its difference for different sectors is hard to determine.

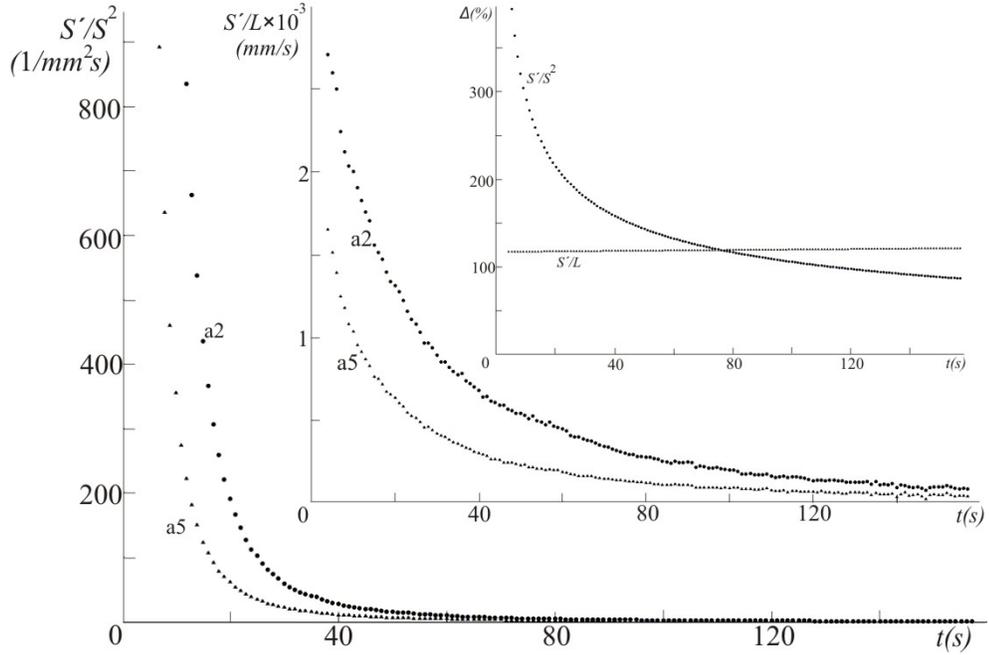

FIG. 8. For the sectors a2 and a5 (Fig. 5), the dependence of the dendrite area increment divided by the squared area ($S'(t)/S^2(t)$) and the dependence of the dendrite area increment divided by the perimeter of the crystal boundary inside the sector ($S'(t)/L(t)$) are given. The insert shows the relative deviation of the values of $S'(t)/S^2(t)$ measured for the sectors a2 and a5, and the relative deviation of the values of $S'(t)/L(t)$ for the same sectors.

ii) What will happen if the sector, where the dendrite area is measured, is directed angle-wise to the growing dendrite branch? Will differences occur here by analogy with the simplest case considered in the Introduction (e.g. sector p2, Fig.1)? The measurements have indicated that the answer is affirmative. As it is shown in Fig. 9, the differences between the values of $S'(t)/S(t)$ sharply increase for different turns of the sector. The error is 30 to 70 percent when turning around 30 degrees, and more than 120 percent when turning around 45 degrees.

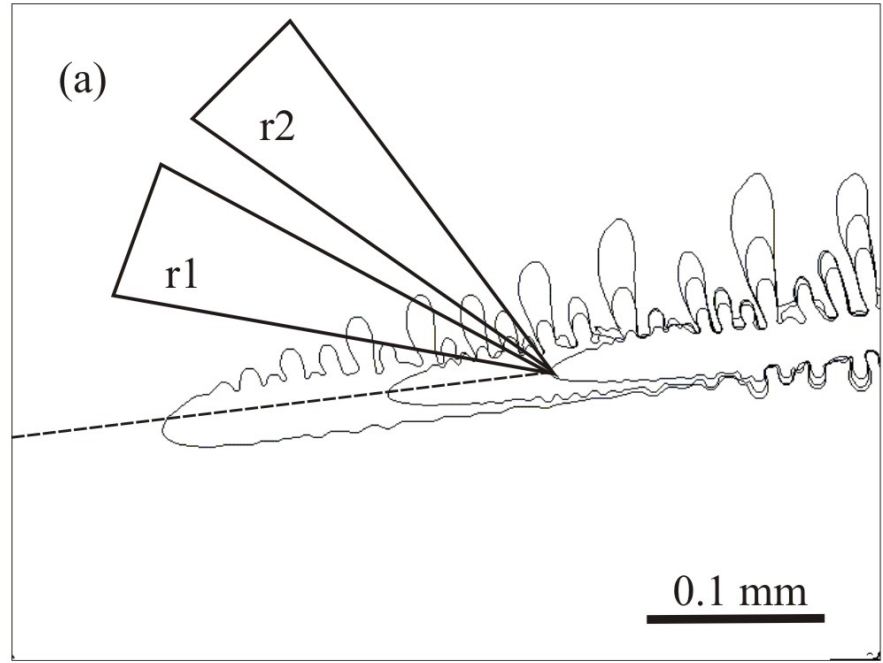

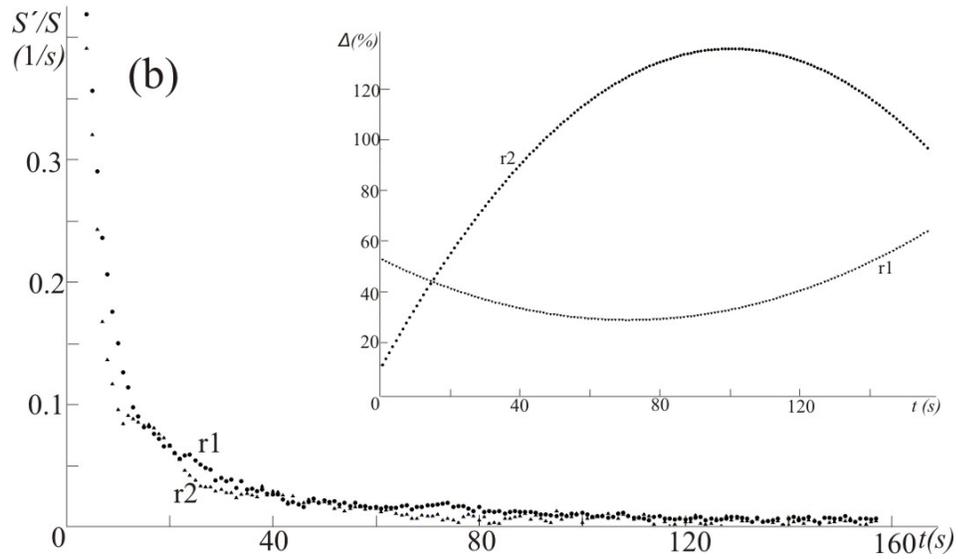

FIG. 9. Values of the normalized area increment $S'(t)/S(t)$ relative to the dendrite growth time $t$ for the sectors (see insert (a)) turned around 30 and 45 degrees with respect to the primary dendrite branch. The insert (b) shows the relative difference between $S'(t)/S(t)$ of each sector and $S'(t)/S(t)$ for the sector a2 Fig.5(a).

Let us refer to a more detailed analysis of the dependence $S'(t)/S(t)$. Table I gives factors of the experimental point approximation using the three-parameter dependence $a/t^c - b$. The correlation coefficient is at least 0.99. According to the table data, $c = 0.97 \pm 0.06$ for the dendrite shown in Fig. 2, and $c = 1.00 \pm 0.04$ for the dendrite in Fig. 3 (with the significance level of 95 percent). Thus, provided the accuracy is high, the factor multiplying the time can be assumed to equal one.

Table I. Values of the three-parameter approximation of experimental data for different sectors.

| Sectors / Factors | a1 | a2 | a3 | a4 | a5 | a6 | a7 | a8 | a9 | a10 | s1 | s2 | s3 | s4 | s5 | s6 | s7 |
|---|---|---|---|---|---|---|---|---|---|---|---|---|---|---|---|---|---|
| a | 1.49 | 1.92 | 1.65 | 1.72 | 1.71 | 1.89 | 1.56 | 1.55 | 1.43 | 1.07 | 1.99 | 1.77 | 1.66 | 1.92 | 1.72 | 1.87 | 1.51 |
| b | 0.02 | 0.01 | 0.01 | 0.01 | 0.01 | 0.00 | 0.01 | 0.01 | 0.01 | 0.02 | 0.02 | 0.02 | 0.01 | 0.01 | 0.01 | 0.01 | 0.02 |
| c | 0.89 | 1.00 | 0.98 | 1.02 | 1.05 | 1.09 | 1.01 | 0.95 | 0.89 | 0.81 | 0.96 | 0.97 | 0.99 | 1.06 | 1.01 | 1.07 | 0.95 |

Table II gives factors of the experimental point approximation using the two-parameter dependence $a/t - b$. The correlation coefficient still proves to be at least 0.99. As it is seen, the factors for different sectors of the same dendrite are close to each other and their distribution is nearly normal. Furthermore, in terms of the experiment accuracy, it is was found that the values of the factor $a$ agree for the sectors of *different* dendrites within the error limits, and the values of the factor $b$ do not agree. Therefore, it is reasonable to mention one average value of the factor $a$, which is $1.7 \pm 0.2$. For the factor $b$, the values, on the average, do not vary for different sectors of one dendrite, but do vary for different dendrites. Thus, $b = 0.007 \pm 0.002$ for the dendrite shown in Fig. 2 and $b = 0.014 \pm 0.002$ for the dendrite in Fig. 3. Let us note that the use of one-parameter approximation of the form $a/t$ proves to be too crude for the available experimental data.

Table II. Values of the two-parameter approximation of experimental data for different sectors.

| Sectors / Factors | a1 | a2 | a3 | a4 | a5 | a6 | a7 | a8 | a9 | a10 | s1 | s2 | s3 | s4 | s5 | s6 | s7 |
|---|---|---|---|---|---|---|---|---|---|---|---|---|---|---|---|---|---|
| a | 1.8 | 1.9 | 1.7 | 1.7 | 1.6 | 1.6 | 1.5 | 1.7 | 1.7 | 1.5 | 2.1 | 1.9 | 1.7 | 1.8 | 1.7 | 1.7 | 1.6 |
| b, ×10⁻² | 0.7 | 1.0 | 0.8 | 1.0 | 0.8 | 0.9 | 0.7 | 0.6 | 0.5 | 0.3 | 1.7 | 1.5 | 1.3 | 1.6 | 1.5 | 1.4 | 1.0 |

If we know that $S'(t)/S(t) = a/t - b$, the analytical dependence for $S(t)$ can be found. The differential equation is solved as follows: $S(t) = C \cdot t^a \cdot \exp(-b \cdot t)$, where $C$ is the constant that can be found using the known value of the crystal area at some moment of time.

The obtained factor $a$ proves to be close to the values obtained analytically when considering the parabola "growth" (see the Introduction). Let us recall that it equals two if the steady growth (kinetically limited regime) is assumed, and equals one for the quasi-steady growth (diffusion-limited regime). The obtained value was between them. The factor $b$ characterizes the decrease of the crystal growth rate[11]. However, this rate decrease is not connected with the geometric features of the growing crystal (as it is case with the diffusion-limited growth in the solution with the constant supersaturation control), but is probably related to the ratio between the initial supersaturation and the time of this supersaturation dropping to zero. Indeed, the initial supersaturation for the dendrite in Fig. 3 and Fig. 2 is approximately the same, however it is reduced almost to zero for around 90 seconds and $b \approx 0.014$ in the first case (Fig. 3), and for around 160 seconds and $b \approx 0.007$ in the second case (Fig. 2). If the supersaturation was maintained constant (as is the case with the examples considered in the Introduction), that would result in infinitely long time of the supersaturation change and $b = 0$. Based on the above, it becomes clear why the factor $b$ (as opposed to the factor $a$) is different for different crystals observed in the experiments. Obviously, this factor will also differ when observing the same crystal at different moments of time. Indeed, we can select sectors not at one moment of time (as it was done in Fig. 5a and Fig. 6a) but at different moments of time. For example, we can place the sector a2 (Fig. 5a) on the top of the primary branch 120 seconds before the crystal growth stops, and then study the behavior of $S'(t)/S(t)$ when the crystal grows through the sector. This sector can also be placed on the top of the primary branch, for example, 70 seconds before the crystal growth stops, and repeat the measurements. In the first case, $b \approx 0.007$, and in the second case, $b \approx 0.004$ (note that the parameter $a$ is still within the interval of $1.7 \pm 0.2$) [12].

---

[11] It is clearly seen from the fact that $S(t) \sim \exp(-b \cdot t)$.

Thus, the brief analysis conducted herein shows that the factor $a$ is of sufficiently universal nature and is apparently connected with the growth regime[13]. As it is known [27], this regime is determined by the ratio between the rate of surface processes (which rate depends on the kinetic factor of crystallization and the characteristic size of the crystal) and the rate of the diffusion transfer in the solution (related to the diffusivity)[14]. The factor $b$ is presumably not a universal characteristic of the crystallized system, it evidently takes into account unsteadiness inherent in the spontaneous (non-controlled) crystallization.

Let us nondimensionalize the obtained dependences $S(t)$ and $S'(t)/S(t)$. The time $t^*$ of the crystal growth termination in the selected sector is the most important time for the problem under discussion. This time can be easily determined from the condition that $S'(t^*) = 0$. It is obvious that

---

[12] In the second case, the supersaturation proves to be considerably lower than in the first case, because the supersaturation decreases exponentially rather than linearly with time.
[13] For the fixed shape of the sector, where the observation is carried out, and the fixed dimension of problem.
[14] Essentially, the growth regime is determined by the basic physical and chemical parameters of the crystallized system [27].

$t^* = a/b$. Let us reduce the dependence of area on time to the dimensionless form, so that $\tilde{S}(t^*) = 1$. As a result, $\tilde{S}(\tilde{t}) = \tilde{t}^a \cdot \exp(a \cdot [1 - \tilde{t}])$ in dimensionless units, where the dimensionless time $\tilde{t} = t/t^*$ and the dimensionless area $\tilde{S} = S / [C(a/b)^a \cdot \exp(-a)]$. Thus, for dimensionless units, all dependences of the crystal area on time obtained in the experiment are described with the one-parameter function $\tilde{t}^{1.7} \cdot \exp(1.7 \cdot [1 - \tilde{t}])$ (Fig. 10)

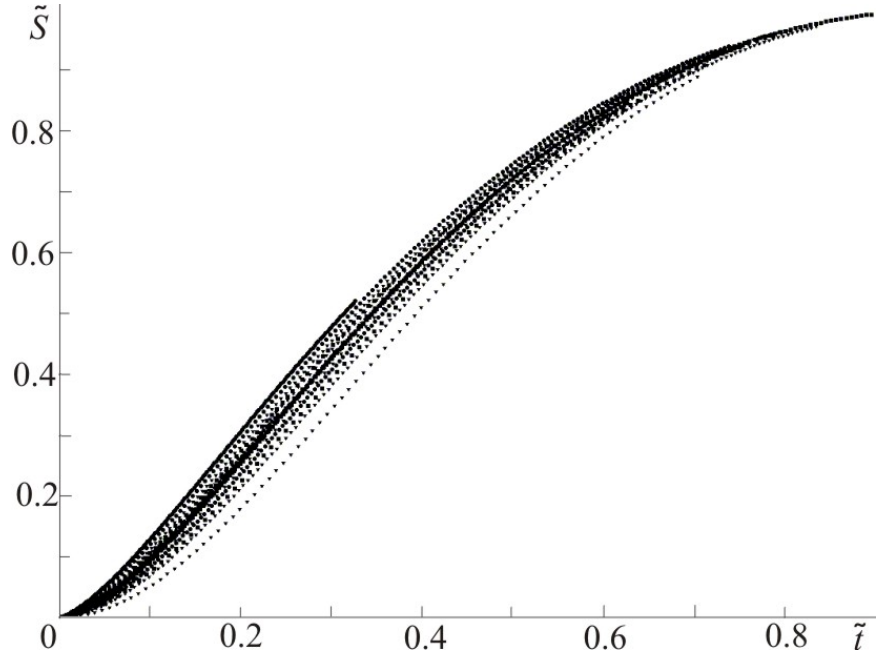

FIG. 10. Values of the nondimensionalized areas of all sectors shown in Fig. 5, Fig. 6 relative to the nondimensionalized time. The nondimensionalization was carried out according to the rules given in the present paper. The parameters $a$ and $b$ for the nondimensionalization were individually taken from Table II for each sector.

If we use the same scale in terms of time, the normalized area increment will have the form $\tilde{S}'/\tilde{S} = a \cdot (1/\tilde{t} - 1)$ or, based on the results of our experiments with the ammonium chloride crystallization, $\tilde{S}'/\tilde{S} = 1.7 \cdot (1/\tilde{t} - 1)$. Figure 11 shows the extent to which the obtained and pre-nondimensionalized data deviate from this universal dependence. For clarity, this graph is constructed using the converted coordinates in which the dependence $\tilde{S}'/\tilde{S} = 1.7 \cdot (1/\tilde{t} - 1)$ has the form of a linear function starting from the origin of coordinates. The presented results (Fig. 10 and Fig. 11) indicate that the obtained experimental data regarding the dependence of the area and the normalized area increment on time can be described well (within 10-15 percent error limits) using quite simple one-parameter dependences after the corresponding normalization.

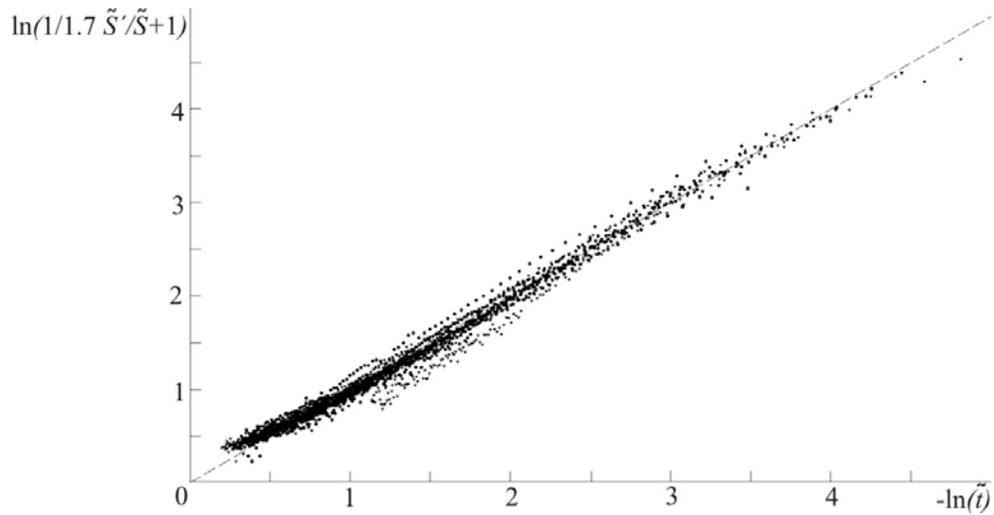

FIG. 11. Values of the nondimensionalized normalized area increment of all sectors shown in Fig. 5, Fig. 6 relative to the nondimensionalized time. The nondimensionalization and the conversion of the coordinate axes was carried out according to the rules given in the present paper. The parameters *a* and *b* for the nondimensionalization were individually taken from Table II for each sector.

## IV. CONCLUSIONS

On the basis of the maximum entropy production principle and the simplest geometrical reasoning, a hypothesis of the universality of the normalized mass increment for the unsteadily growing dendrite is advanced herein. This hypothesis is confirmed in the present paper based on the experiment with the quasi-two-dimensional crystallization of ammonium chloride from an aqueous solution. It is found that the dendrite area change divided by the area itself varies with time as $a/t - b$ for any sector oriented along the growth direction of the dendrite branches (primary or side). The values of the factors *a* and *b* are obtained and their physical meaning is discussed. This result allows proposing an analytical form of the curve for describing the evolution of the dendrite area (or its part) with time of the form $S(t) = const \cdot t^a \cdot \exp(-b \cdot t)$ [15]. The nondimensionalization of quantities using the full dendrite growth time enables to reduce S(t) and $S'(t)/S(t)$ to simple one-parameter dependences in which the parameter is presumably connected with the dendrite growth regime in the solution. The given results are the first study so far to find a universal (invariant) characteristic for the unsteadily growing (primary and side) dendrite branches. Additional experiments with both ammonium chloride and other dendrite-forming substances are required for complete certainty in the

---

[15] There is no definite opinion regarding that issue in the literature. Particularly, the Weibull function is used to describe the dependence of the crystal mass (area) on time [17,26,27,38]

obtained results and to progress in the selected direction of study[16].

---

[16] Apparently, the obtained result for the area (in the case of the quasi-two-dimensional system) can be generalized to the crystal volume (for the three-dimensional case). In that case, the values of the factors will certainly change.